\documentclass[letterpaper, 12pt, fleqn]{article}
\usepackage[margin=1.0in]{geometry}
\usepackage[linkcolor=black, urlcolor=black, colorlinks=true, citecolor=black]{hyperref}
\usepackage[utf8]{inputenc}
\usepackage[english]{babel}

\usepackage{ebgaramond-maths}
\usepackage[T1]{fontenc}

  \usepackage{
	enumerate,
	fancyhdr,
	ifthen,
	hanging,
	setspace,
	lastpage,
	pdfsync,
	natbib,
	amsfonts,
	xcolor,
	newpxmath,
	xspace,
	array,
	siunitx
	}

\bibpunct{(}{)}{:}{:}{,}{,}
\setcitestyle{notesep={: }}

\newlength\mystoreparindent

\ifpdf
	\usepackage[pdftex]{graphicx}
	\DeclareGraphicsExtensions{.pdf, .jpg, .tif}
\else
	\usepackage{graphicx}
	\DeclareGraphicsExtensions{.eps, .jpg}
\fi
\begin{document}

\title{How (and Why) to Think that the Brain is Literally a Computer
\vspace{0.25in}

{\Large forthcoming in \emph{Frontiers in Computer Science, Section: Theoretical Computer Science}. 
(doi:\href{https://doi.org/10.3389/fcomp.2022.970396}{10.3389/fcomp.2022.970396}})
}

\author{Corey J. Maley\\cmaley@ku.edu} 

\maketitle

\begin{abstract}
The relationship between brains and computers is often taken to be merely metaphorical. However, genuine computational systems can be implemented in virtually any media; thus, one \emph{can} take seriously the view that brains literally compute. But without empirical criteria for what makes a physical system genuinely a computational one, computation remains a matter of perspective, especially for natural systems (e.g., brains) that were not explicitly designed and engineered to be computers. Considerations from real examples of physical computers---both analog and digital, contemporary and historical---make clear what those empirical criteria must be. Finally, applying those criteria to the brain shows how we can view the brain as a computer (probably an analog one at that), which, in turn, illuminates how that claim is both informative and falsifiable.

\end{abstract}

\section{Introduction} 
\label{sec:introduction}
The fact that neuroscientists and cognitive scientists have likened the mind and brain to a computer is hardly news. This so-called metaphor is so well-known---and so widely discussed---that some researchers would like to see this metaphor and its attendant debates put to rest \citep{brette2022, richards2022, kelty-stephen2022, gomez-marin2022}. The reasons for wanting to move on from what may seem like a mere metaphor are understandable: there are many ways that one might understand the term ``computer,'' all of which seem reasonable within a particular context, and none of which is to be preferred over any other. Most scientists might agree that they would rather spend their time doing science than worrying about debates over the meanings of words. However, philosophers of computation and neuroscience are more than happy to worry about the meaning of words; not because it is interesting for its own sake, but because it can illuminate the assumptions we bring to the table when we try to think carefully and communicate effectively about the brain. This is especially important in interdisciplinary areas like cognitive science, computer science, neuroscience, where interrogating our concepts can be a fruitful way to allow for researchers from different areas to come together, and avoid talking past one another.

Here, it will be argued that the brain may well be a computer in a literal sense, not just metaphorically. This requires empirical criteria for what makes a \emph{physical} system a \emph{computational} system. Such criteria go beyond the conceptual and mathematical resources of theoretical computer science; a candidate set of criteria will be developed here.

With these criteria in hand, it will be shown how the brain can be a literal computer, although probably not a \emph{digital} computer. The account of physical computation sketched here shows both what makes different species of computation different (e.g., digital, analog, and perhaps others) and what makes them all count as \emph{bona fide} computational types.

Taken together, these considerations show both that the brain can literally engage in computation, and that whether it does so is an empirical claim. Moreover, although analog computation is a bona fide type of computation, it is different from digital computation in non-obvious ways relevant to neuroscientists, cognitive scientists, and theoretical computer scientists. A few of those differences will be discussed.


\section{Computational Theory and Its Limits} 
\label{sec:computational_theory_and_computational_criteria}

The foundations of computation are often understood to begin with the mathematical work of Alan \citet{turing1936}. The standard view regarding what it is for a physical object to \emph{be} a computer is that the object implements a Turing Machine (TM) or some other abstract automaton. Many interesting questions about the nature and limits of computation have come from the theory of computation (TOC) and theoretical computer science,\footnote{I will abbreviate all of theoretical computer science as TOC in what follows, just for convenience.} including many limitative and complexity results.

As valuable as TOC is, we must also be clear about what it does not (and cannot) do. In particular, TOC does not have the theoretical resources to provide criteria for what makes a \emph{physical} object a computer. To be sure, once we have decided that some physical object \emph{is} a computer, then TOC can tell us much about the efficiency and limitations of the computations that object can perform. But TOC has no criteria for making the initial decision about which physical objects compute in the first place, and which do not.

Of course, for most computational systems, we already know they are computers, simply because we have designed and engineered them as such. The decision has already been made. But faced with a natural object such as a brain, we need criteria for making such a decision, and the TOC is of no help.

An analogy is useful. Number theory does not have the resources to tell us how to count physical objects, or even which physical objects can be counted at all (e.g., some physical things, like water, can only be counted when unitized). If given a photograph of clouds, for example, one might wonder whether the number of clouds is prime. But how do we individuate clouds? It is an obvious mistake to look to number theory for the answer: one must look elsewhere, perhaps to meteorology. However, once we have decided on criteria to count clouds, then we can use the resources of number theory to answer questions about numbers of clouds. While admittedly a simple (and even simplistic) example, the point is the same: we must look beyond the TOC for criteria of what counts as a computer.

At this point, it is worth revisiting an important fact about physical computational systems. All computational systems are what some have called ``medium independent'' or ``substrate neural'' or ``multiply realizable,'' meaning that they can be implemented in virtually \emph{any} physical system with the right properties.\footnote{There are some technical differences between these terms, but for our purposes, they do not matter.} Obviously there are practical benefits to using electronic circuitry for the computers we use every day, but theoretically, computers can be (and have been) implemented with mechanical gears, fluid flow, marbles, and Tinkertoys. A computer made of such seemingly exotic materials is not merely a computer in a metaphorical sense: it is \emph{literally} a computer, just as much as a computer made of silicon or gallium arsenide. Similarly, a neural implementation of a computer would not be a computer in some metaphorical sense: it would \emph{literally} be a computational system.

This all depends on empirical criteria for what makes a physical system computational in the first place, as just mentioned. Let us look at a proposal for just such a set of criteria.

\section{Empirical Criteria for Physical Computation} 
\label{sec:empirical_criteria_for_physical_computation}
As important as Turing's work is, it was \emph{not} the conceptual birthplace of computation.\footnote{A complete story would also discuss the work of Church and G\"{o}del, among others, but that is for a longer essay.} It is somewhat well-known (but not as well-known as it should be) that Turing modeled his \emph{a}-machines (what we now call Turing Machines) on the actions of \emph{human} computers. These people (most often women) were employed to solve mathematical problems for scientific, research, and engineering purposes \citep{ceruzzi1991, light1999}.

Moreover, Turing was only interesting in modeling \emph{some} of the actions of human computers. One type of computation involves manipulating numbers digitally. Much like the algorithms for addition or long-division learned by many school children, this type of computation involves representing numbers by strings of digits, then performing operations on those digits. The resulting digit-string is the output of the computation. However, human computers also used \emph{analog} devices, such as slide rules. These devices do \emph{not} represent numbers by their digits, but by their \emph{magnitudes}. 
\begin{figure}[!ht]
	\begin{center}
	\includegraphics[width=4.0in]{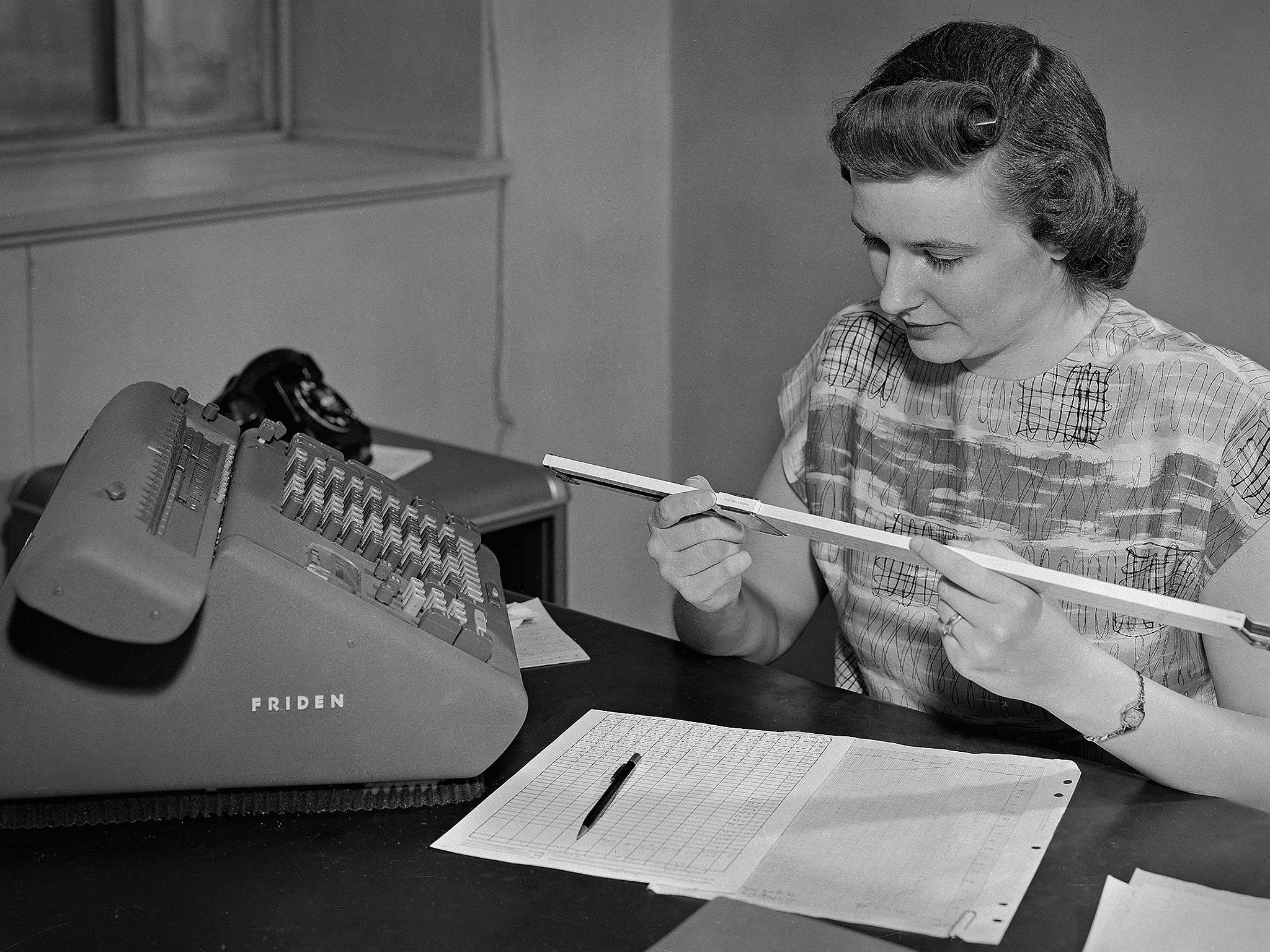}
	\end{center}
	\caption{Human computer at the National Advisory Committee for Aeronautics (NACA) Lewis Flight Propulsion Laboratory using a slide rule and adding machine.}\label{fig:1}
\end{figure}
Analog computing mechanisms are simply not amenable to analysis by Turing Machines or any other digital automaton: they just do not represent numbers in the right way.\footnote{Of course, digital computers can approximate analog computations, but the very fact that they \emph{approximate} demonstrates that the way these computations are performed is fundamentally different.} For Turing's purposes, this was perfectly acceptable. However, it does illustrate that computation both predates Turing's theoretical work, and that Turing's theoretical work does not capture---and was never intended to capture---every type of computation. Both analog and digital computing devices have existed for millennia before Turing's analysis.

So, computation and computing mechanisms existed before Turing, and computation has never been limited only to the digital. What is it that unites both analog and digital computation, as well as uniting what human computers did (i.e., the basis of Turing's mathematical analysis \emph{of} computation) and what computing mechanisms did (and still do)?

On a first pass, the answer seems simple: computation is the manipulation of numbers (i.e., one or more inputs) in order to produce another number as output. But it is not quite that simple; we cannot manipulate numbers directly\footnote{Whatever it is that numbers are, they are not spatiotemporal objects---mathematicians need not do empirical studies to discover the properties of mathematical objects.} so we have to manipulate concrete \emph{representations} of numbers. Even that is not quite good enough though, because not just any manipulation counts. The numerals ``86'' written on a chalkboard can be manipulated by erasing the top half of each numeral, leaving two symbols that look a bit like ``oo,'' but this is clearly not a computation.

The \emph{right} kind of manipulation is difficult to pin down, but it helps to compare the kinds of manipulations that \emph{do} count when we use computing devices. Consider a digital computer, where the binary representation of 86 would be 01010110. In the circuit element where this number is represented, each of the 1s would be represented by (say) five volts, and each of the 0s by zero volts. A manipulation that counts as a computation would involve manipulating those voltages; heating those elements with a hair dryer, or painting them red, are manipulations that would not count. So we have to manipulate voltage, which is precisely the property that is doing the representing.

If we consider an analog electronic computer, we find a similar thing, although the type of representation works a bit differently. In this case, the number 86 would be represented by a single circuit element at 86 volts. Again, a manipulation that counts as a computation would involve manipulating that voltage, and not some other property. In short then, we can say that a computation has to involve the manipulation of a representation, where the manipulation is \emph{of the very property that is doing the representing}, and not some other property incidental to the relevant representational capacities.

Finally, when it comes to computing devices, the manipulation has to be performed by a mechanism that is sensitive only to the properties just mentioned. A significant amount of literature in the philosophy of science has been devoted to articulating precisely what is meant by a ``mechanism,'' but for now, we need not go into these details (a good reference point is \citep{piccinini2007c}).

We can now put everything together: \emph{something is a physical computing device when it manipulates physical representations via a mechanism, where that mechanism is sensitive to, and only manipulates, the physical properties of representations that do the representing.}

If this articulation of a physical computing device is correct---that is, if it does justice to what we already take to be computers, and have taken to be computers in the past, both analog and digital---then we can use it as an empirical criterion for what it means for \emph{anything} to be a physical computing device: not just the ones we have designed as such. And indeed, it does just that: computing devices, from abaci and the Antikythera mechanism to slide rules and contemporary digital computers all do their computing work via the mechanistic manipulation of representations, where the manipulations are of just the properties responsible for doing their representing.

Now, before moving on, it is worth addressing an objection, or what may seem an oversight. What counts as computation in theoretical computer science often involves the activities of abstract automata which manipulate symbols that do not represent anything at all. Any decent undergraduate course in the theory of computation involves showing that certain classes of automata (PDAs), but not others (DFAs), can recognize languages like $a^n b^n$, where the a's and b's are uninterpreted, discrete symbols that do not represent anything. The concern is with the manipulation of symbols that \emph{could} be representations, but usually are not. This seems to be at odds with the characterization of computation just offered, where representation is central.

There is much to say on this point, and philosophers of computation continue to debate this issue (\citep{piccinini2015} and \citep{shagrir2022} are both excellent monographs on opposing sides). For the sake of brevity, I will simply say that the theoretical/mathematical notion of computability that is the focus of the theory of computation (when physically implemented) is simply a superset of the types of manipulations that count as physical computations. After all, a great many physical mechanisms implement a great many automata of the type just mentioned: automata that do not manipulate representations, but uninterpreted symbols. In fact, nearly \emph{every} physical mechanism with discrete states implements some such automaton. Elevators, light switches, gear-shifting mechanisms on bicycles, deadbolt locks\ldots the list goes on. However, we reserve the term ``computer'' for physical mechanisms that manipulate representations.\footnote{To be sure, there are some who view the entire universe as one giant computing mechanism, but I set that view aside here: if everything is literally a computer, then the notion that the brain, or anything else, is a computer, is trivial.} Moreover, if we return to the human computers modeled by Turing Machines, a person manipulating digits in the right way, or the lengths of a slide rule, would be computing. But a person manipulating uninterpreted symbols (e.g., moving a's and b's around, even according to a well-defined rule, where the a's and b's are not the digits of numbers or anything meaningful at all) would not be computing anything, and neither would a person just moving a couple of sliding sticks back and forth, where their lengths do not represent anything.
\section{Neural Computation} 
\label{sec:neural_computation}
Now that we have an outline of empirical criteria for when a physical system is computational, we can apply these criteria to determine whether brains do, in fact, compute. This amounts to asking whether the brain uses representations, and if so, whether the properties of those representations are manipulated by mechanisms that are only sensitive to the properties that do the representing.

On both counts, the answer is a clear ``yes.'' The idea that neural systems traffic in representations is nearly orthodox in the cognitive sciences. Moreover, the dominant strategy for explaining a neuroscientific phenomenon of interest is to identify and articulate the mechanism responsible for that phenomenon \citep{craver2007}. Combining the identification of representations with the identification of mechanisms that manipulate them just is what many neuroscientists already do when reporting the results of neural computation.

Let us look at a simple, yet classic, example. \citet{roeder1966, roeder1998} describes, in fascinating detail, the neural computation involved in the noctuid moth's attempts to escape the echolocating bats that prey upon them. In short, the tympanic organ of the moth detects sound waves, generating neural spikes with a frequency that increases as the intensity of the sound increases. When the sound intensity is low, moths steer away from the source of the sound; when the sound intensity becomes high enough, the moths engage in rapid, erratic behaviors such as closing their wings completely. In other words, when bats are far away, moths can attempt to simply fly away from them; but when these predators are very close, the moths must engage in quick, evasive maneuvers.

In this case, the frequency of neural spikes represents the relative closeness of the moth (or relative intensity of sound). Neural circuitry then uses that information---specifically, the neural spike frequency---to decide between different behaviors, driving the muscles of the wings.\footnote{There are more details, of course, but they demonstrate the point even further.} Thus, we have both a representation and a neural mechanism that manipulates that representation, which then generates adaptive behaviors. I have argued elsewhere \citep{maley2011, maley2018} that this would be a paradigm analog representation, because one magnitude (neural spike frequency) is a representation of another magnitude (decibel level). In any case, it is clearly a representation. Coupled with a mechanism that manipulates that representation, we have neural computation.

The benefit of adopting the empirical criteria presented here is that whether a neural system performs computations is not a matter of perspective or semantics. If the system contains representations, and if they are manipulated in the right way, then it simply is computational. Now, disagreements about what counts as a representation in the first place may arise between different researchers: although the term is ubiquitous in neuroscience, there is room for more precision in what different neuroscientists mean by the term \citep{baker2021a, cao2022}. On the other hand, there are clearly many aspects of neural function that are \emph{not} representational; like every other living cell, neurons do a vast number of things that having nothing to do with processing information. The criteria advocated here make explicit what many neuroscientists agree with in the first place: not everything the brain does is computational.

Finally, this account of physical computation explains what is different about different types of computation. Digital computation is digital because it uses digital representations; analog because it uses analog. Here again, we need clear accounts of what we mean by these terms. I have noted elsewhere that, both conceptually and as a matter of historical fact, analog computers can be discrete; they represent the magnitudes of variables via the magnitudes of physical quantities, whereas digital computers represent the digits of variables via differences in physical quantities (which need not be magnitudes) \citep{maleyforthcominga, maleyforthcomingb}. What makes different types of computation different is the type of representation used, which, in turn, constrains the types of mechanisms that can manipulate those representations. Perhaps neural computations are analog, because neural representations are analog (as I have argued elsewhere); perhaps they are something else entirely. In any case, what makes each of these species of the genera \emph{computation} is that they all manipulate representations via mechanisms of the right kind.

\section{Discussion} 
\label{sec:discussion}
Given the empirical criteria for physical computation developed above, the claim that the brain is a computer is both meaningful and falsifiable. It is not a matter of mere stipulation. If the brain does not use representations, or if it does, but those representations are not manipulated by mechanisms that are sensitive \emph{only} to the properties responsible for doing the representing, then the brain (or whatever relevant part of it) does \emph{not} compute. Providing empirical criteria for determining whether the brain computes, using a principled characterization of computation, is an improvement upon much of the recent discussion. Even more though, according to this account, it seems to be true that the brain does compute.

Many other ideas follow from a better understanding of computation in general, and how neurons might actually be engaged in non-digital computation, such as analog computation. Some have claimed that computation requires software, or the ability to be programmed \citep{brette2022}; others have claimed that computation requires universality of some type \citep{richards2022}. But analog computation, for example, is different in both of these respects, and yet still counts as computation. Moreover, analog computation may not even admit of a distinction between what \citet{marr1982} called the algorithmic/representational level and the implementational level \citep{maley2021}. We may have taken what is true \emph{only} of digital computation, or \emph{only} of computation based on the implementation of Turing Machines (and the like), to be true of \emph{all} computation. And while it would be a mistake to advocate for a return to analog computation in general, we should attend to \emph{all} computational types when it comes to thinking about how natural systems might compute.

\end{document}